# Early Prediction of Post-acute Care Discharge Disposition Using Predictive Analytics: Preventing Delays Caused by Prior Health Insurance Authorization and Reducing the Inpatient Length of Stay

**Avishek Choudhury**
Stevens Institute of Technology
USA

Correspondence:

**Name**: Avishek Choudhury

**Qualification**

Systems Engineering (Ph.D. Student, Stevens Institute of Technology)

Applied Data Science (MS, Syracuse University)

Industrial Engineering (MS, Texas Tech University)

**Contact**:

achoudh7@stevens.edu

achoud02@syr.edu

avishekadamas@gmail.com

Mobile # +1 (515) 608-0777

**ORCID:** https://orcid.org/0000-0002-5342-0709

Researcher ID: P-2415-2018

**Website:** https://avishekchoudhury1.academia.edu

Word count – 3291 (excluding references and title page)

**Acknowledgments** The authors thankfully acknowledge the procedural and clinical assistance of Mrs. Shelby Neel (Nurse, UnityPoint Health) and Ms. Vanessa Calderon (Process Improvement Engineer, UnityPoint Health)

**Competing interests statement** No conflict of interest

**Funding:** This study was not funded by any internal or external sources.

**Ethics approval:** All data were collected with the permission of the organization (UnityPoint Health, Des Moines), and there was no disclosure of any of the medical or personal information of the patients.

**Data statement:** The raw anonymized data and data descriptions are available at http://dx.doi.org/10.17632/ghk2df4cvc.1. Refer to the DOI for data citations and citing this article.

# Early Prediction of Post-acute Care Discharge Disposition Using Predictive Analytics: Preventing Delays Caused by Prior Health Insurance Authorization and Reducing the Inpatient Length of Stay

**ABSTRACT**

*Objective:* A patient's medical insurance coverage plays an essential role in determining the post-acute care (PAC) discharge disposition. The prior health insurance authorization process postpones the PAC discharge disposition, increases the inpatient length of stay, and effects patient health. Our study implements predictive analytics for the early prediction of the PAC discharge disposition to reduce the deferments caused by prior health insurance authorization, the inpatient length of stay and inpatient stay expenses. *Methodology:* We conducted a group discussion involving 25 patient care facilitators (PCFs) and two registered nurses (RNs) and retrieved 1600 patient data records from the initial nursing assessment and discharge notes to conduct a retrospective analysis of PAC discharge dispositions using predictive analytics. *Results:* The chi-squared automatic interaction detector (CHAID) algorithm enabled the early prediction of the PAC discharge disposition, accelerated the prior health insurance process, decreased the inpatient length of stay by an average of 22.22%, and reduced inpatient stay expenses by $1,974 for state government hospitals, $2,346 for non-profit hospitals and $1,798 for for-profit hospitals per day. The CHAID algorithm produced an overall accuracy of 84.16% and an area under the receiver operating characteristic (ROC) curve value of 0.81. *Conclusion:* The early prediction of PAC discharge dispositions can condense the PAC deferment caused by the prior health insurance authorization process and simultaneously minimize the inpatient length of stay and related expenses incurred by the hospital.

## 1. INTRODUCTION

Post-acute care (PAC) has involved multiple providers administering aid in a disconnected manner and poor communication throughout the health care system (Abrams, O'Rourke, & Gerhardt, n.d.). When a patient requires PAC services, there is currently little reason given as to why a patient is discharged to an SNF, a home health agency, an AR facility, or a long-term acute care hospital (Burrill, n.d.). Delayed PAC can result in poor care, higher than average readmission rates, and suboptimal patient outcomes (Burrill, n.d.).

The growth of the elderly patient population may cause a variety of medical challenges in healthcare industries. Elderly patients tend to require lengthy hospital stays and post-acute care (PAC) assistance to attain desirable health restitution (Yu-Shan, Chu-Sheng, Yao-Hsien, & et.al, 2012) (McKee & J, 2001). More than one-third of stroke patients in the United States are discharged to PAC facilities, including acute rehabilitation (AR), skilled nursing (SNFs) and long-term care facilities (W, 2016). The need for PAC has been growing since 2001. One out of five patients is admitted to PAC after being discharged from the hospital (about 8 million patients annually) (Tian, 2016). On an average, 22.8 % of SNF patients end up back in the hospital within 30 days of their discharge (Burke, et al., 2015).

PAC requires prior health insurance authorization (Cigna-HealthSpring , 2017). In 2014, patients suffering from neurological diseases comprised 13% of Medicare cases in AR, up from 5% in 2004 (A data book: healthcare spending and the medicare program). This increase led to an increase in Medicare spending, which grew from $20.3 billion in 2001 to $41.3 billion in 2014 (A data book: healthcare spending and the medicare program).

A prior health insurance authorization delays PAC services (Mills, 2018) and increases the inpatient length of stay. Moreover, prior health authorization issues are concomitant with 92% of care deferments, and they contribute to patient wellbeing issues and administrative ineptitudes (Mills, 2018). According to an AMA survey that assessed the experiences of a thousand patient care physicians, 64% reported delays for prior authorization decisions from insurers of at least one business day, and 30% stated they wait three to four business days or longer (Mills, 2018). In addition, 8 out of 10 physicians said the hinderances related to prior health insurance authorization were high, and 86% of these physicians believed that burdens associated with prior authorization have increased over the past five years (Mills, 2018) and led to increases in Medicare spending, PAC obligations and the services provided by insurance companies (MB, et al., 2005), including bundled SNF payments.

The inpatient stays of patients discharged to PAC are typically lengthier and more expensive than routine discharges (Yu-Shan, Chu-Sheng, Yao-Hsien, & et.al, 2012) (McKee & J, 2001). The stay length and cost are influenced by the complexity of medical conditions (TM, Baker DI, & Peduzzi PN, 2002) (LK, et al., 2010) (FI & DW, 1965) and PAC facility placement delays caused by prior health insurance authorization requirements. The Institute for Healthcare Improvement says that hospital-wide patient flow should deliver “the right care, in the right place, at the right time” (MP & EM, 1969). PAC discharge dispositions require the meticulous coordination of insurance administrators and admitted patients (R, S, & K, 2006). PAC discharge dispositions are also affected by the availability of required settings, the accessibility of the patient, and pecuniary incentives that might not be allied with medical requirements or cost effectiveness (H, R, & R, 2007). A whitepaper from the Institute for Healthcare Improvement suggests working with AR and SNF facilities to improve patient flow through advanced planning, coordination and partnership development (MP & EM, 1969). However, no significant research has been performed to address advanced PAC discharge disposition planning and improved coordination between acute and post-acute services (JA, et al., 1982).

PAC services, when provided on time, have improved the physical independence and recovery of patients (A data book: healthcare spending and the medicare program) (MB, et al., 2005). In addition to programs in the US, a PAC program that began in 2006 in Taiwan improved the bodily activities, psychological function, and pain management of patients.

In this study, we propose an advanced PAC discharge disposition plan using predictive analytics and address the delay caused by prior health insurance authorization, which is responsible for increased inpatient lengths of stay, deferred PAC services, and increased inpatient stay expenses. Advanced PAC discharge disposition planning identifies patients eligible for post-acute services (AR or SNF) based on their initial nursing assessment early in their inpatient stay, enabling the prior health insurance authorization process to start earlier.

## 2. METHODOLOGY

This study does not involve patient participation, and no personal patient information are revealed. All analysis and patient data were anonymized for legal and ethical purposes.

The methodology of this study can be broadly categorized into the following sections: (a) group discussion and problem identification, (b) data collection, (c) data preprocessing, and (d) model selection and assessment.

### 2.1. **Group discussion and problem identification**

To study the PAC discharge disposition procedures and determine the bottlenecks responsible for PAC discharge delays and long inpatient stays, we conducted a group discussion involving 25 patient care facilitators (PCFs) and two registered nurses (RNs).

The three main questions that were discussed in this session are as follows:

- What criteria do we use to determine whether a patient should go to acute rehabilitation?

- What criteria do we use to determine whether a patient should go to a skilled nursing facility?
- Is there a defined process map that we follow before a patient discharge note is signed by a doctor? (only to the RNs)

In this section, we collected all the information necessary for understanding and mapping the existing process so that the bottleneck region could be identified.

## 2.2. Data collection

We retrieved 1600 patient data records (from July 2018 through August 2018) from discharge and preoperative assessment notes. The data consist of 306 attributes, including 1 response variable. The raw anonymized data (Choudhury, 2018) are used for all analyses. Our analysis included only patients discharged to AR or SNF, and missing data and deceased patient data were excluded. All data were available in the EPIC database. Figure 2 shows the EPIC database interface used at a non-profit hospital in Iowa.

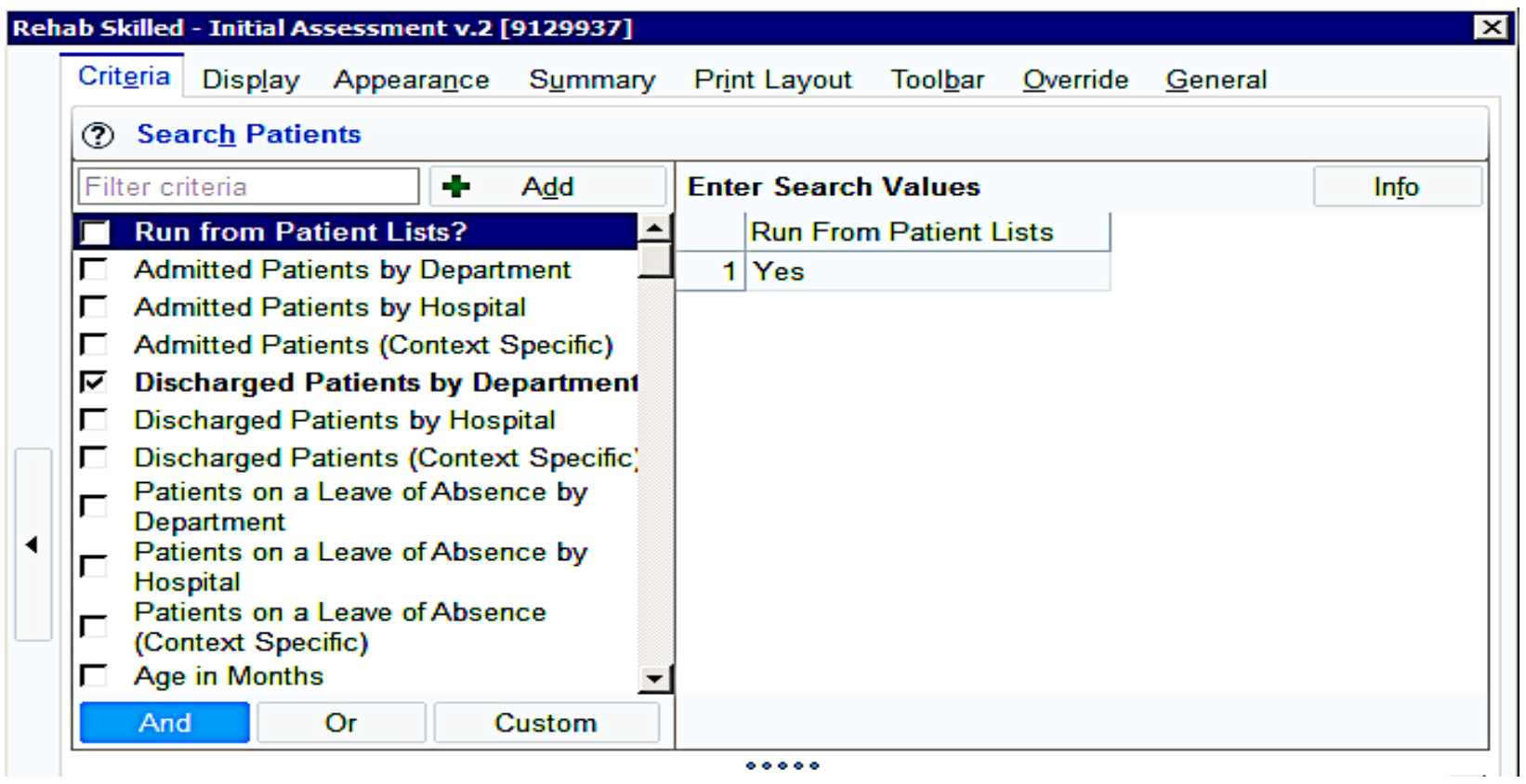

**Figure 1:** EPIC database interface used by non-profit hospital, Iowa. All patient information was retrieved from this secure platform.

## 2.3. Data description

In this section, we briefly describe the data and illustrate the continuous predictors (Hester-Davis fall risk score and Braden scale score) and their behaviors with respect to the age, gender and discharge disposition of the patient.

**Table 1:** Discharge disposition and gender

| Discharge Disposition | Gender | | |
|---|---|---|---|
| | **Male** | **Female** | **Missing** |
| Another Health Care Institution Not Defined | 2 (0.13%) | | 0 |
| Federal Hospital | 4 (0.26%) | | 0 |
| Psychiatric Hospital | 5 (0.33%) | | 0 |
| Rehab Facility | 24 (1.58%) | 14 (0.92%) | 0 |
| Short-term General Hospital for Inpatient Care | 4 (0.26%) | 2 (0.13%) | 0 |
| Skilled Nursing Facility | 76 (5.01%) | 114 (7.52%) | 0 |
| Swing Bed | 1 (0.06%) | 1 (0.06%) | 0 |
| Intermediate Care Facility | 12 (0.79%) | 17 (1.12%) | 0 |
| Home Health Care Service | 75 (4.95%) | 45 (2.97%) | 0 |
| Long-term Care | | 3 (0.19% | 0 |
| Expired | 13 (0.85%) | 8 (0.52%) | 0 |
| Home or Self Care | 499 (32.93%) | 552 (36.43%) | 0 |
| Hospice | 7 (0.46%) | 5 (0.33%) | 1 |
| Hospice Medical Facility | 11 (0.72%) | 6 (0.39%) | 0 |
| Left Against Medical Advice | 10 (0.66%) | 3 (0.19%) | 0 |
| Court/Law Enforcement | 1 (0.06%) | | 0 |

**Table 2:** Discharge disposition, Braden scale score, Hester-Davis fall risk and patient age

| **Discharge Disposition** | **Average Braden Scale Score** | **Average of Hester-Davis Fall Risk Score** | **Average Age** |
|---|---|---|---|
| Another Health Care Institution Not Defined | 20 | 7 | 64 |
| Federal Hospital | 13 | 12 | 68 |
| Psychiatric Hospital | 15 | 9 | 49 |
| Rehab Facility | 17 | 11 | 66 |
| Short-term General Hospital for Inpatient Care | 17 | 9 | 59 |
| Skilled Nursing Facility | 16 | 12 | 76 |
| Swing Bed | 15 | 15 | 92 |
| Intermediate Care Facility | 15 | 14 | 73 |
| Home Health Care Service | 18 | 9 | 65 |
| Long-term Care | 15 | 12 | 79 |
| Expired | N/A | N/A | N/A |
| Home or Self Care | 20 | 7 | 57 |
| Hospice | 18 | 14 | 72 |
| Hospice Medical Facility | 16 | 13 | 79 |
| Left Against Medical Advice | 19 | 7 | 51 |
| Court/Law Enforcement | 15 | 26 | 40 |

**Table 3:** Descriptive statistics

| **Predictor** | **Age** | **Braden Scale Score** | **Hester-Davis Fall Risk Score** |
|---|---|---|---|
| Min | 16 | 1 | 3 |
| Max | 97 | 26 | 23 |
| Range | 81 | 25 | 20 |
| Mean | 71.90 | 12.44 | 15.52 |
| Mean Std. Error | 1.37 | 0.42 | 0.37 |
| Std. Deviation | 15.56 | 4.83 | 4.18 |
| Variance | 242.16 | 23.39 | 17.52 |
| Skewness | -1.30 | 0.15 | -1.01 |
| Skewness Std. Error | 0.21 | 0.21 | 0.21 |
| Kurtosis | 2.51 | -0.16 | 1.21 |
| Kurtosis Std. Error | 0.42 | 0.42 | 0.42 |

Figures 3 and 4 show the age distribution by discharge type (AR and SNF) and by sex, respectively. Figures 5 and 6 show the relationship between age and the Braden scale score and between age and the Hester-Davis fall risk score, respectively.

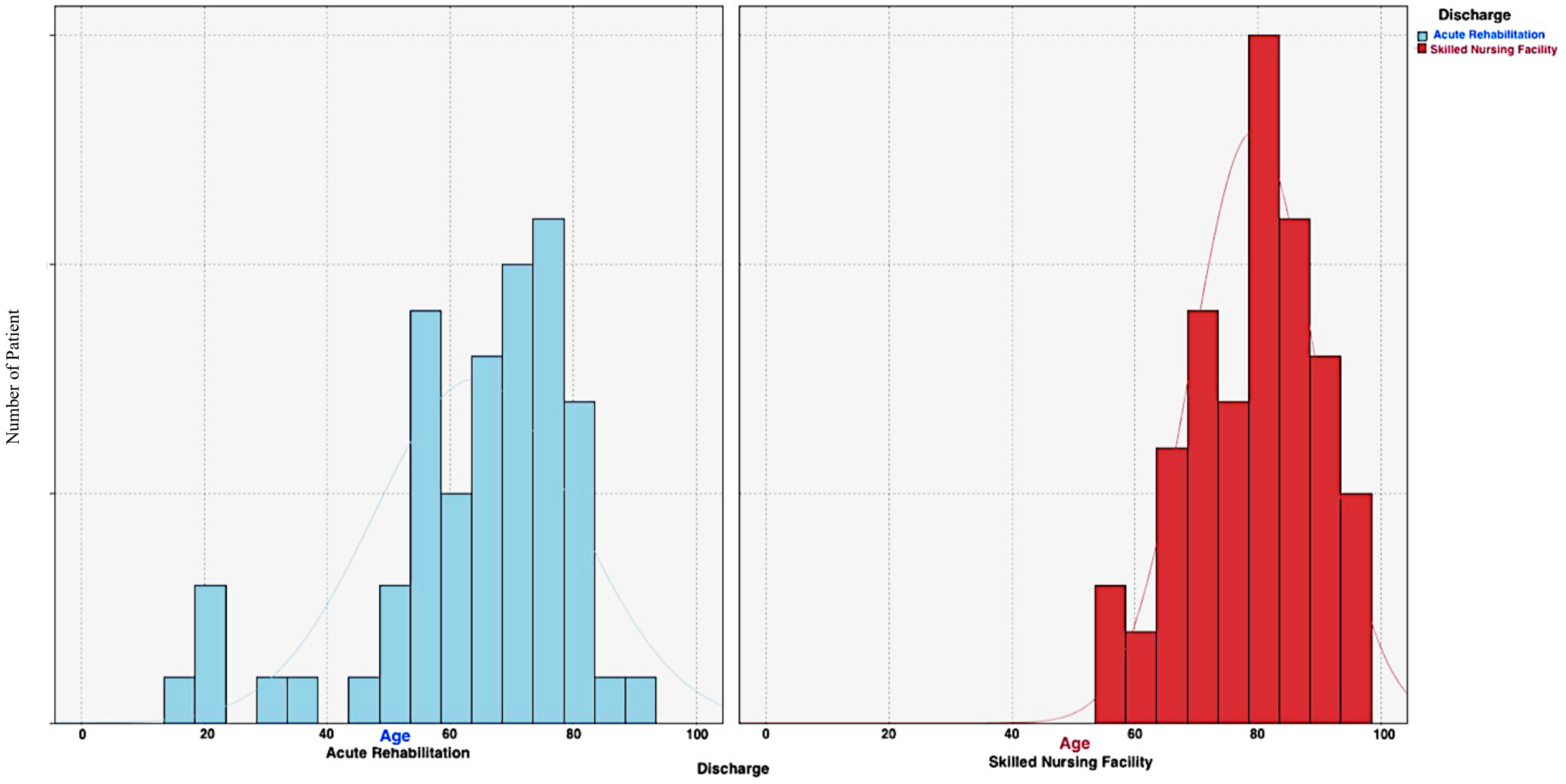

**Figure 2:** Age distribution by discharge type. The blue histogram illustrates the age of patients discharged to AR, and the red histogram shows the age of patients discharged to SNF. Older patients were more likely to be discharged to SNF, and younger patients were more likely to tolerate the required three hours of therapy.

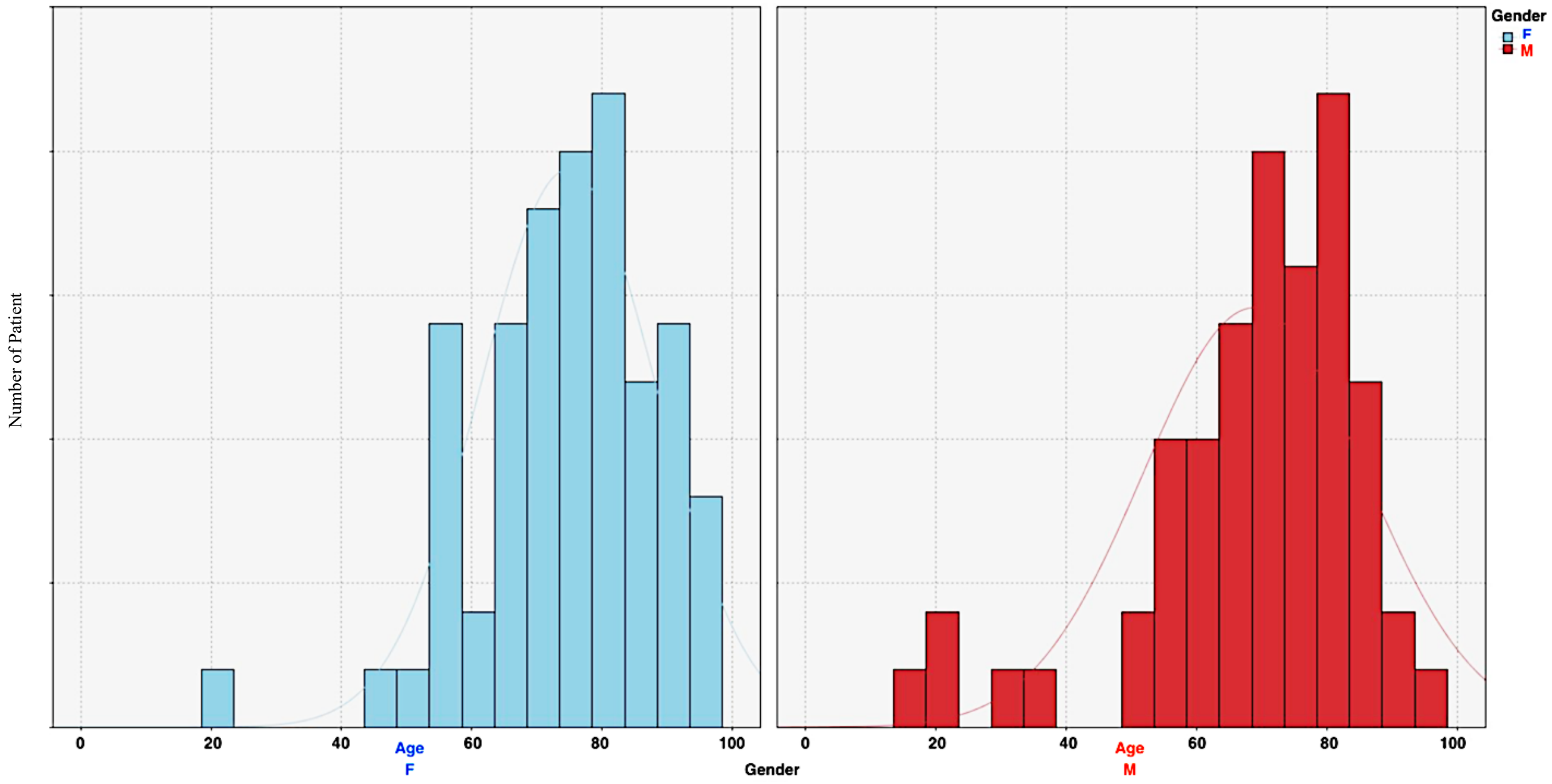

***Figure 3:*** *Age distribution by gender.* The blue histogram shows the age distribution of female patients, and the red histogram shows the age distribution of male patients.

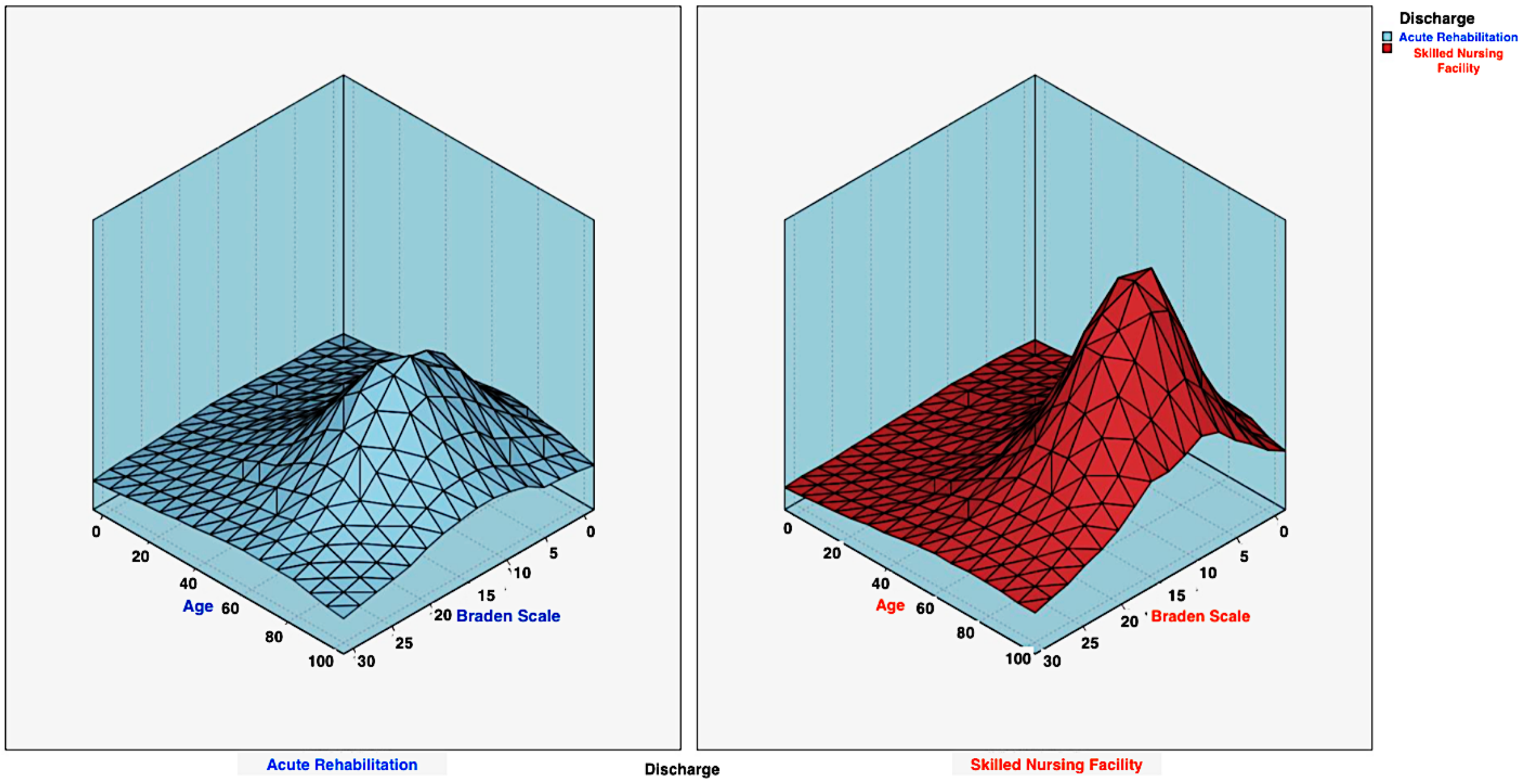

**Figure 4:** The relationship between patient age and the Braden scale score. The blue surface shows the Braden scale score of AR patients, and the red surface shows the Braden scale score of SNF patients. The patients discharged to SNF have higher Braden scale scores than the patients discharged to AR.

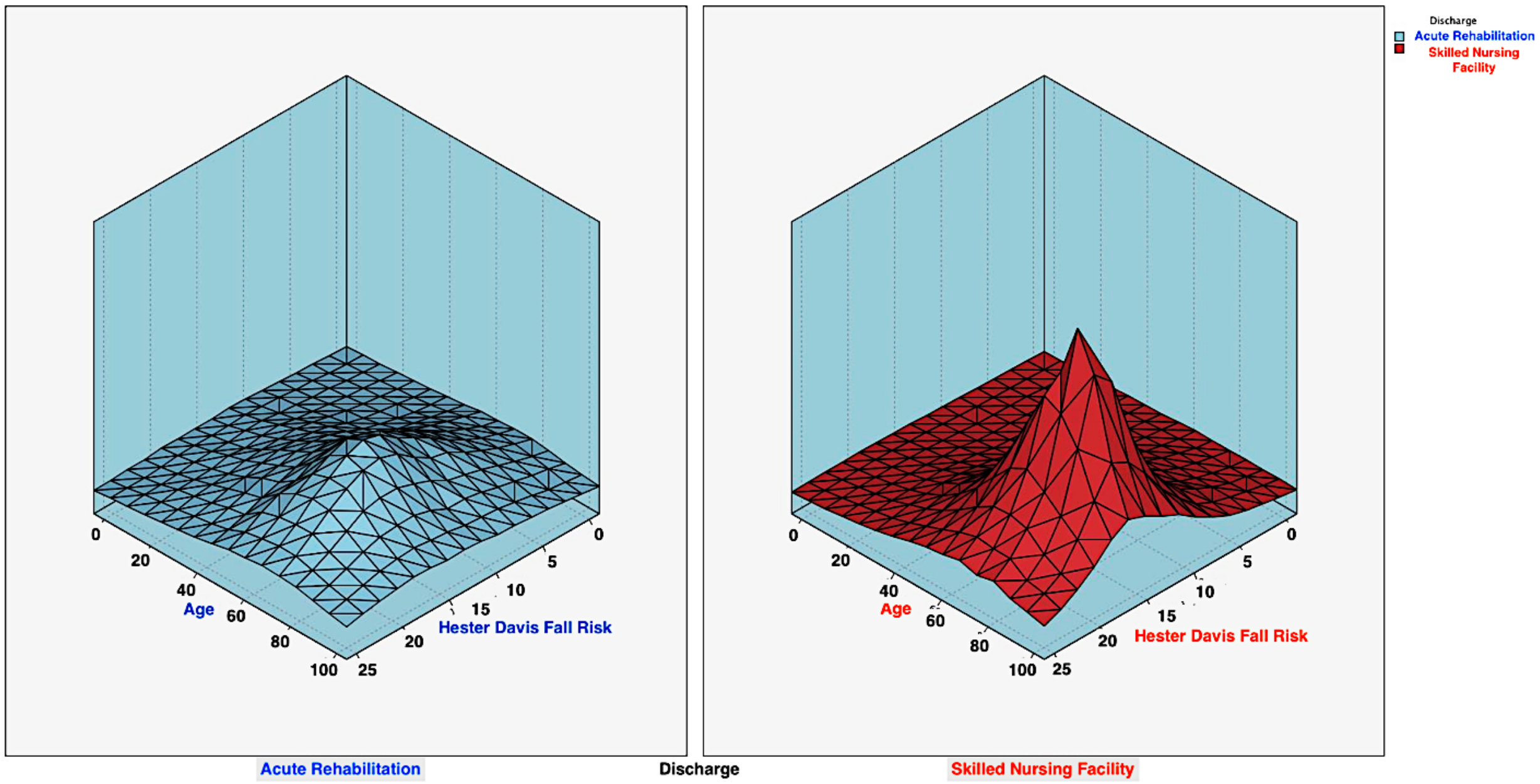

**Figure 5:** The relationship between the Hester-Davis fall risk and patient age. The blue surface shows the Hester-Davis fall risk of AR patients, and the red surface shows the Hester-Davis fall risk of SNF patients. SNF patients have a higher Hester-Davis fall risk than AR patients.

### 2.4. Data preprocessing

The retrieved data set was divided into training (70%) and testing (30%) records.

### Model selection

We implemented the following five machine learning algorithms: **(a)** linear discriminant analysis (LDA) (Brownlee, Linear Discriminant Analysis for Machine Learning, 2016), **(b)** the chi-squared automatic interaction detector (CHAID) (Marley), **(c)** a random tree (RT) method (Rapidminer, n.d.), **(d)** a linear support vector machine (LSVM) (Brownlee, Machine learning mastery, 2016), and **(e)** a classification and regression tree (CART) (Loh, 2011). The model that provided the best fit was chosen based on the following three performance measures: (a) overall accuracy, (b) lift above 30%, and (c) area under the ROC curve.

## 3. RESULTS

### 3.1. Group discussion of outcomes

## Early commencement of prior health insurance authorization

The 25 PCFs and 2 RNs agreed that the PAC discharge type is primarily driven by medical insurance coverage and the physical therapy (PT) and occupational therapy (OT) evaluations of a patient. They also noted the importance of health conditions, such as a stroke, hip fracture or spinal cord injury, in mandating AR service for a patient. However, to qualify for AR, the patient must be able to tolerate three hours of therapy and must be covered by medical insurance.

Figure 6 shows an approximate protocol that PCFs, RNs and doctors typically follow to manage PAC discharges. The figure was developed based on information gathered from the group discussion and includes only the crucial steps involved in actual practice that are relevant to this study.

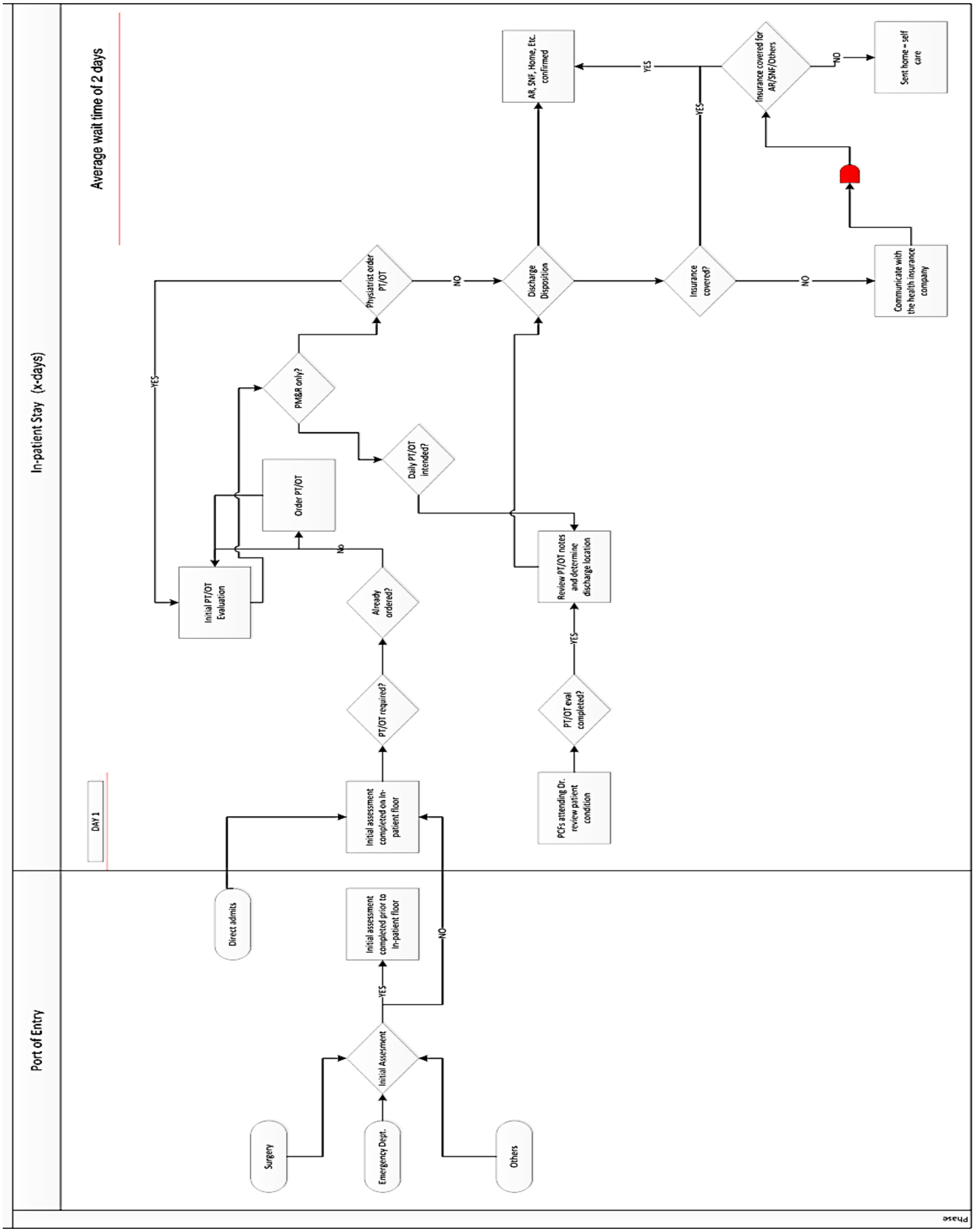

**Figure 6:** The existing PAC discharge disposition process (for a traditional practice without predictive modeling). In this practice, the hospital requests insurance converge after all clinical procedures are completed. Then, the patient and the hospital wait for two days on an average for the insurance coverage confirmation. This two-day waiting time adds no value to the healthcare services of the patient but increases the inpatient length of stay, hinders patient health and delays PAC (a high-resolution picture is provided).

### Early commencement of prior health insurance authorization

After a discharge decision is made and confirmed by both the PCF and the doctor, the hospital initiates the prior health insurance authorization process; it takes two days on average *(Beaton, 2018)* for the insurance company to confirm whether a patient is insured for AR or SNF, thus postponing the discharge by two days. This process was identified as the bottleneck region responsible for PAC discharge disposition delays and long inpatient stays.

## 3.2. **Comparative analysis of predictive models**

Several machine learning algorithms were used to predict which patients are eligible for AR and SNF services early in the preoperative phase. The CHAID algorithm, with the highest overall accuracy of 84.16% and a ROC value of 0.81, was selected as best fit model. Table 5 shows the accuracy and the area under the ROC curve of the top five tested models.

**Table 4:** Comparative analysis of predictive models

| Sl. No. | Model | Overall Accuracy | Area Under the ROC |
|---|---|---|---|
| **1** | LDA | 83.33 | 0.79 |
| 2 | CHAID | 84.16 | 0.81 |
| **3** | RT | 72.50 | 0.68 |
| **4** | LSVM | 76.66 | 0.70 |
| **5** | CART | 80.00 | 0.51 |

## 3.3. **Advantage and impact of the CHAID predictive model**

Implementing the CHAID algorithm yielded the following results: (a) a reduction in the inpatient length of stay, (b) timely PAC discharge, and (c) reduced inpatient stay expenses.

A predicted PAC discharge disposition can be determined on the first day of an inpatient stay, and the average length of an inpatient stay decreased from *x* days to *x-2* days. Table 6 shows the reduction in the average inpatient length of stay due to the early prediction of the PAC discharge disposition.

**Table 5:** Comparison of PAC discharge times using the traditional method and the CHAID algorithm (predictive modeling)

| **PAC Discharge Method** | **PAC Service (average time)** | **Prior Authorization (average time)** | **Total Time (average)** |
|---|---|---|---|
| Traditional Method | *X* days | Additional 2 days | *X+2* days |
| After Implementing CHAID | ***X* days (including the prior authorization time)**<br>*(PAC service: day 1 through day X)*<br>*Medical insurance confirmation (day 1 through day 2)* | 2 days | *X* days |
| Total Reduction in Time (average) | | | **2 days** *(22.22%)* |

Figure 7 shows the process map after implementing CHAID. The CHAID model can identify eligible AR and NSF patients during the initial nursing assessment, thereby enabling the hospital to initiate the prior health insurance authorization process on the first day of an inpatient stay (rather than at the end of the inpatient stay). Thus, PT/OT evaluation, initial continued nursing assessment, and all other essential clinical activities can be processed while the medical insurance company confirms the patient's insurance coverage, and the patient will not have to wait an extra 2 days to obtain health insurance authorization after the doctor recommends the discharge location. Moreover, this model does not interfere with clinical processes or replace physician decisions. The model is designed to encourage and enable advanced PAC discharge disposition planning by proactively gauging medical insurance coverage in parallel with the inpatient stay. The new process map ensures the recursive training of the CHAID model, which enhances its reliability and robustness over time and provides a support system for all medical experts.

Early commencement of prior health insurance authorization

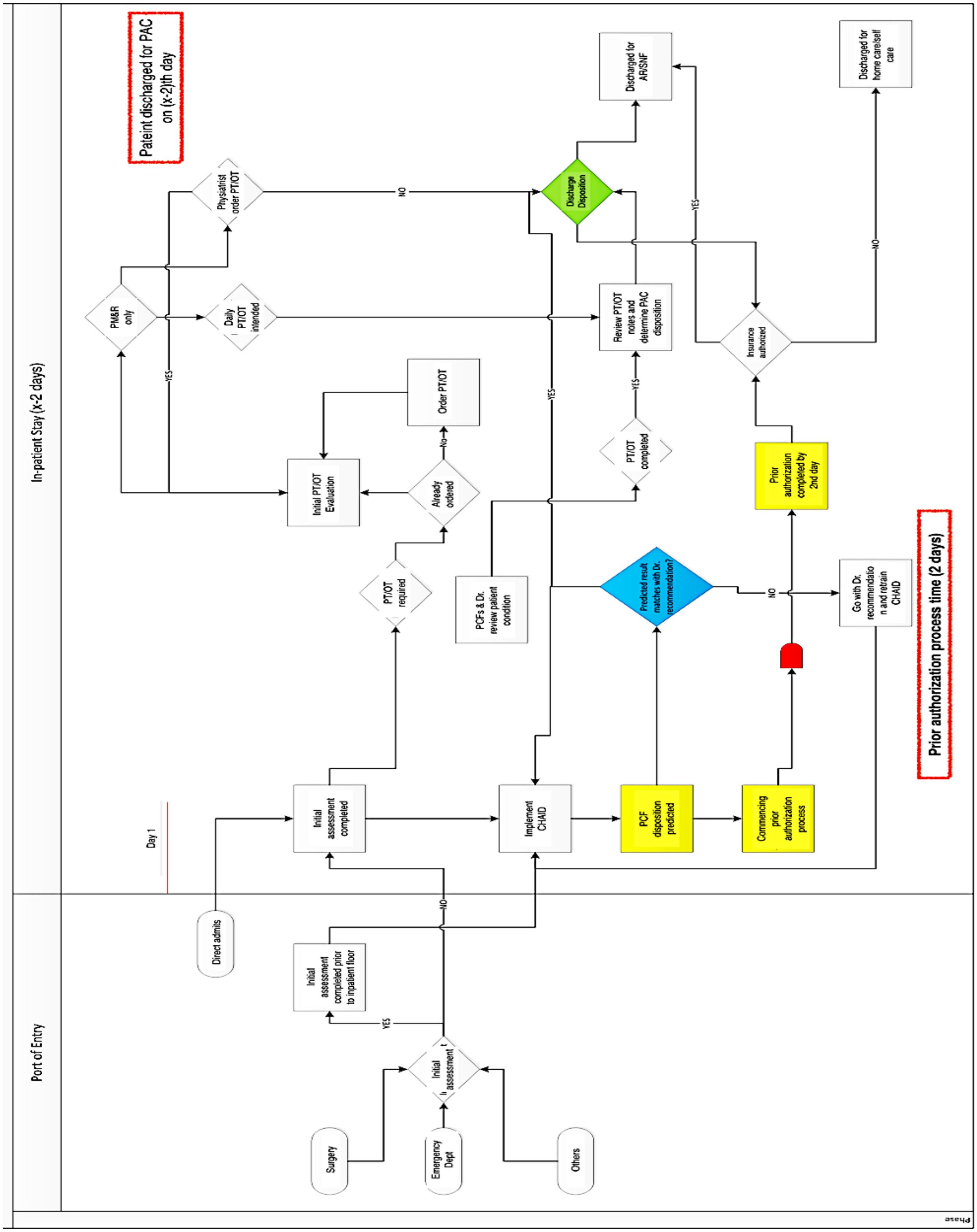

**Figure 7:** Process map showing the advantage of implementing the CHAID model. The CHAID model removes the extra two days of waiting time so that prior health insurance authorization can be performed in parallel with the other clinical procedures conducted during an inpatient stay, thereby reducing inpatient length of stay (a high-resolution picture is provided).

The proposed model reduces both the inpatient length of stay and the expenses per inpatient day. The cost of an inpatient stay has rapidly increased, and the table below shows the trend for all operating and non-operating costs in registered US community hospitals (defined as non-federal short-term general hospitals within the United States of America) (1999 - 2016 AHA Annual Survey, n.d.).

**Table 6:** Hospital adjusted expenses per inpatient day

| Year (1999-2016) | Inpatient Stay Expenses/ Day/Patient Incurred by the Hospital ($) | Change in Expenses (%) (A) | Moving Average (A) |
|---|---|---|---|
| **1999** | 1102 | | |
| **2000** | 1148 | 4.06 | |
| **2001** | 1217 | 5.65 | |
| **2002** | 1290 | 5.63 | 5.11 |
| **2003** | 1371 | 5.91 | 5.73 |
| **2004** | 1450 | 5.48 | 5.67 |
| **2005** | 1522 | 4.73 | 5.37 |
| **2006** | 1612 | 5.57 | 5.26 |
| **2007** | 1696 | 4.94 | 5.08 |
| **2008** | 1782 | 4.84 | 5.12 |
| **2009** | 1853 | 3.80 | 4.53 |
| **2010** | 1910 | 2.99 | 3.87 |
| **2011** | 1960 | 2.55 | 3.11 |
| **2012** | 2090 | 6.24 | 3.93 |
| **2013** | 2157 | 3.10 | 3.96 |
| **2014** | 2212 | 2.50 | 3.95 |
| **2015** | 2271 | 2.57 | 2.72 |
| **2016** | 2338 | 2.88 | 2.65 |

Note: The adjusted expenses per inpatient day include expenses incurred for both inpatient and outpatient care. It is important to note that these data are only an estimate of the expenses incurred by the hospital to provide a day of inpatient care and are not a substitute for either actual charges or reimbursement for care provided.

The following figure shows the increase in the cost of an inpatient stay from 1999 through 2016.

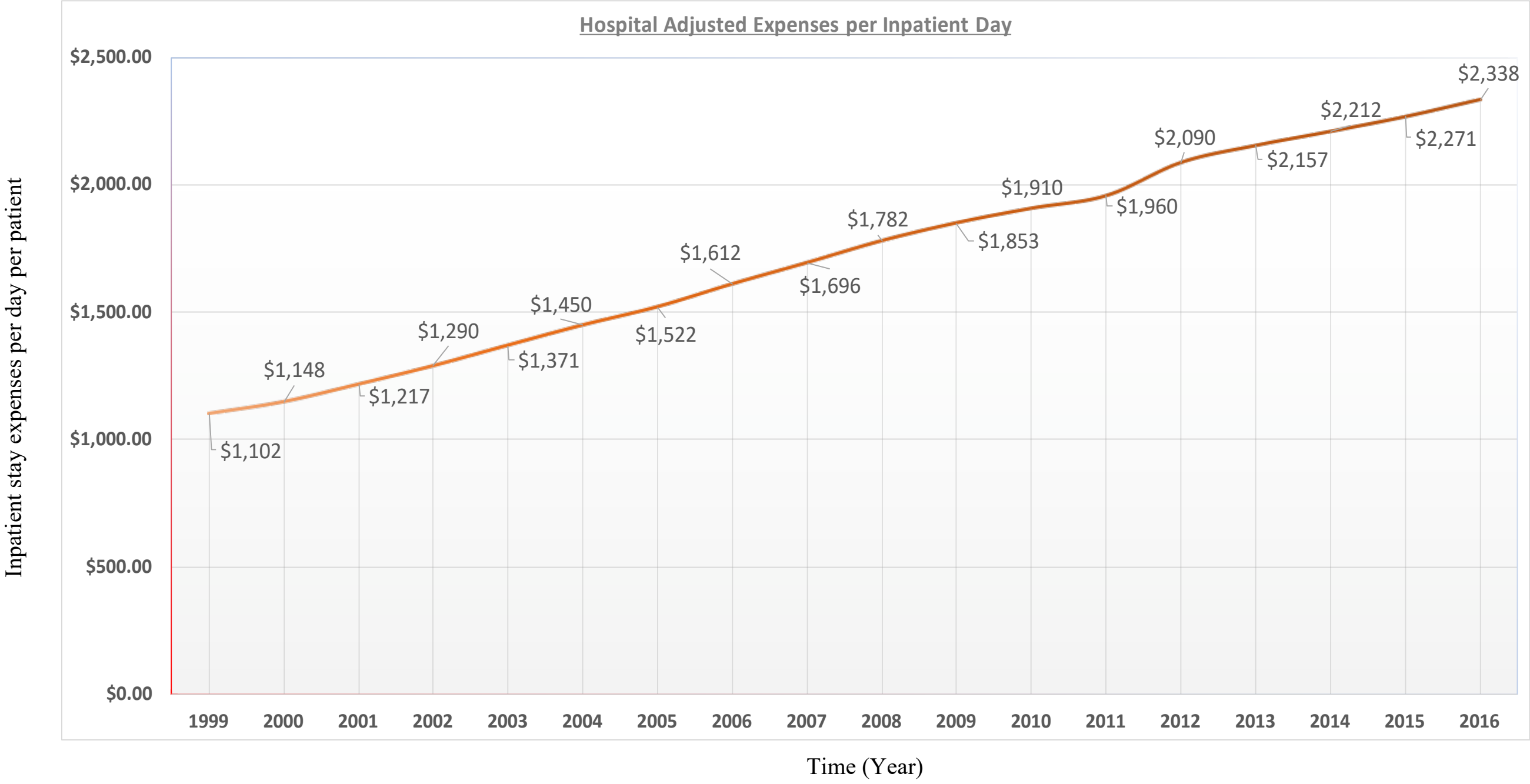

**Figure 8**: Increasing trend of inpatient stay expenses per person within the United States of America

On average, the prediction of the PAC discharge type and the early commencement of prior health insurance authorization reduces the inpatient length of stay by 2 days and reduces the cost per inpatient day. The early prediction of the PAC discharge disposition can yield cost reductions of $1,974 for state government hospitals, $2,346 for non-profit hospitals and, $1,798 for for-profit hospitals per day (Ellison, 2016).

## 4. CONCLUSIONS

Our study is the first that implements predictive modeling to determine the PAC discharge type in advance. CHAID is the best fit model, with an overall accuracy of 84.16%. This model has the potential to reduce the inpatient length of stay by 22.22% by encouraging advanced PAC discharge disposition planning, as suggested by the Institute for Healthcare Improvement (MP & EM, 1969). Moreover, implementing the repeatable and reproducible data-driven CHAID model can reduce the inpatient cost of stay by an average of $1,974 for state government hospitals, $2,346 for non-profit hospitals and, $1,798 for for-profit hospitals per day.

Currently, 6-Clicks, which was developed by the Cleveland Clinic Health System (Diane u. Jette, 2014), is an effective and user-friendly tool for assisting in PAC discharge disposition planning (Diane U.Jette, 2014). The 6-Clicks tool, also known as AM-PAC, is a corroborative tool based on the activity domain of the World Health Organization's International Classification of Functioning, Disability and Health (ICF) (Internatinal Classification of Functioning, Disability and Health, 2001); however, it does not focus on advanced PAC discharge disposition planning. The proposed CHAID model outperforms the accuracy of 6-Clicks and is free of user bias.

## 5. SUMMARY

### Findings

1. Healthcare providers and patients are expected to experience lengthy waits before their prior health insurance authorization applications are sanctioned.
2. Seventy-eight percent of providers stated that long prior health insurance authorization processes are associated with patients stopping their treatments.
3. Lengthy inpatient wait times contribute to increased healthcare expenses and poor health outcomes.

### Our contributions

1. This study is the first that implements advanced planning for PAC discharge types to minimize the inpatient length of stay based on predictive analytics.
2. The CHAID algorithm is implemented and yields an accuracy of 84.16%.

3. The study uses real data in the analyses.
4. The PAC discharge time is reduced by 22.22% and the inpatient cost of stay is decreased by an average of $1,974 for state government hospitals, $2,346 for non-profit hospitals and, $1,798 for for-profit hospitals per day based on the proposed model.

**AUTHORSHIP STATEMENT**

# Early Prediction of Post-acute Care Discharge Disposition Using Predictive Analytics: Preventing Delays Caused by Prior Health Insurance Authorization and Reducing the Inpatient Length of Stay

All persons who meet authorship criteria are listed as authors, and all authors certify that they have participated sufficiently in the work to take public responsibility for the content, including participation in the concept, design, analysis, writing, or revision of the manuscript. Furthermore, each author certifies that this material or similar material has not been and will not be submitted to or published in any other publication.

**Authorship contributions**

- Conception and design of study: Avishek Choudhury;
- acquisition of data: Avishek Choudhury;
- analysis and/or interpretation of data: Avishek Choudhury;
- Drafting the manuscript: Avishek Choudhury;
- Figure preparation: Avishek Choudhury;
- revising the manuscript critically for important intellectual content: Avishek Choudhury;